\documentstyle[aps,preprint]{revtex}
\newcommand{\beq}{\begin{equation}}
\newcommand{\eeq}{\end{equation}}
\begin{document}
\draft
\baselineskip=16pt
\tightenlines

\title{Model of Strongly Correlated 2D Fermi Liquids Based
on Fermion-Condensation Quantum Phase Transition}

\author{V.R. Shaginyan \footnote{E--mail:
vrshag@thd.pnpi.spb.ru}} \address{ Petersburg Nuclear Physics
Institute, Russian Academy of Sciences, Gatchina, 188300, Russia}
\maketitle

\begin{abstract}
A model of strongly correlated electron or hole liquids with
the fermion condensate is presented and
applied to the consideration of quasiparticle excitations in high
temperature superconductors, in their superconducting and normal
states. Within our model the appearance of the fermion condensate
presents a quantum phase transition, that separates the regions of
normal and strongly correlated electron liquids.  Beyond the
fermion condensation quantum phase transition point the quasiparticle
system is divided into two subsystems, one containing normal
quasiparticles and the other --- fermion condensate localized at the
Fermi surface. In the superconducting state the
quasiparticle dispersion in systems with FC can be presented by two
straight lines, characterized by effective masses $M^*_{FC}$ and
$M^*_L$, respectively, and intersecting near the binding energy which
is of the order of the superconducting gap. This same quasiparticle
picture persists in the normal state, thus manifesting itself over
a wide range of temperatures as new energy scales. Arguments are
presented that fermion systems with FC have features of a ``quantum
protectorate". A theory of high temperature superconductivity based
on the combination of the fermion-condensation quantum phase
transition and the conventional theory of superconductivity is
presented. This theory describes maximum values of the
superconducting gap which can be as big as $\Delta_1\sim
0.1\varepsilon_F$, with $\varepsilon_F$ being the Fermi level. We
show that the critical temperature $2T_c\simeq\Delta_1$. If there
exists the pseudogap above $T_c$ then $2T^*\simeq\Delta_1$, and $T^*$
is the temperature at which the pseudogap vanishes. A discontinuity
in the specific heat at $T_c$ is calculated. The transition from
conventional superconductors to high-$T_c$ ones as a function of the
doping level is investigated.  The single-particle excitations and
their lineshape are also considered.  Analyzing experimental data on
the high temperature superconductivity in different materials induced
by field-effect doping, we show that all these facts can be
understood within a theory of the superconductivity based on the
fermion condensation quantum phase transition, which can be conceived
of as a universal cause of the superconductivity.  The main features
of a room-temperature superconductor are considered.

\end{abstract}


\section {INTRODUCTION}

One of the most challenging problems of modern physics is the problem
of systems with a big coupling constant. This problem is of crucial
importance particularly in the quantum field theory, making even the
quantum electrodynamics to be not a self-consistent theory \cite{rf}.
It is well-known that a consideration of strongly correlated liquids
is close to the problem of systems with the big coupling constant
which persists in many-body physics. A solution of this problem has
been offered in the context of the Landau theory of normal Fermi
liquids by introducing into the theory a notion of the quasiparticles
and parameters which characterize the effective interaction between
the quasiparticles \cite{lan}.  As a result, the Landau theory has
removed high energy degrees of freedom and kept a sufficiently large
number of relevant low energy degrees of freedom to treat liquid's
low energy properties. Usually, it is assumed that the breakdown of
the Landau theory is defined by the Pomeranchuk stability conditions
and occurs when the Landau amplitudes being negative reach its
critical value. Note that the new phase at which the stability
conditions are restored can in principle be again described in the
framework of the theory. To describe a strongly correlated electron
liquid, taking place when the coupling constant becomes sufficiently
big, a conventional way can be used, assuming that the correlated
regime is connected to the noninteracting Fermi gas by adiabatic
continuity in the same way as in the framework of the Landau normal
Fermi liquid theory \cite{lan}. But a question exists whether this
is possible at all.  Most likely, the answer is negative. Therefore,
we direct our attention to a model, in the frame of which a strongly
correlated electron liquid is separated from conventional Fermi
liquid by a phase transition related to the onset of the fermion
condensate (FC) \cite{ks,vol}. It was demonstrated rather recently
\cite{ks} that the Pomeranchuk conditions are covering not all
possible limitations: one is overlooked, being connected with the
situation when, at temperature $T=0$, the effective mass can become
infinitely big. It has been demonstrated that such a situation,
leading to profound consequences, can take place when the
corresponding amplitude being positive reaches the critical value,
producing a completely new class of strongly correlated Fermi liquids
with FC \cite{ks,vol} which is separated
from a normal Fermi liquid by the fermion condensation quantum phase
transition (FCQPT) \cite{ms}. In this case we are dealing with the
strong coupling limit where an absolutely reliable answer cannot be
given, based on pure theoretical first principle ground. Therefore,
the only way to verify that FC occurs is to consider experimental
facts which bear witness to the existence of such a state. We assume
that these facts can be find in two-dimensional (2D) systems with
interacting electrons or holes, which can be presented by
modulation doped quantum wells, by high mobility
metal-oxide-semiconductor field effect transistors, or by high-$T_c$
superconductors.

The aim of our report is to show that within the
frameworks of our model of a strongly correlated electron
(hole) liquid based on FCQPT the main properties of such liquids
observed in the high-temperature superconductors can be understood.
In Sec. II, we review the general features of Fermi liquids with
FC, showing that an electron liquid of low density  in the
high-$T_c$ materials inevitably undergoes FCQPT.  In Sec. III we
consider the high-temperature superconductivity, which takes place in
the presence of FC. In Sec. IV we describe the quasiparticle
dispersion and lineshape.  Finally, in Sec. V, we summarize our main
results.

\section {THE GENERAL FEATURES OF FERMI LIQUIDS WITH FC}

Let us start by considering the key points of the FC theory. FC is
a new solution of the Fermi liquid theory equations\cite{lan} for
the quasiparticle occupation numbers $n(p,T)$
\beq \frac{\delta(F-\mu N)}{\delta
n(p,T)}=\varepsilon(p,T)-\mu(T)-T\ln\frac{1-n(p,T)}{n(p,T)}=0,\eeq
which depends on the momentum $p$ and temperature $T$. Here $F$ is
the free energy, and $\mu$ is the chemical potential, while
\beq\varepsilon(p,T)=\frac{\delta E[n(p)]}{\delta n(p,T)},\eeq is
the quasiparticle energy. This energy is a functional of $n(p,T)$
just like the total energy $E[n(p)]$ and the other thermodynamic
functions. Eq. (1) is usually presented as the Fermi-Dirac
distribution, \beq
n(p,T)=\left\{1+\exp\left[\frac{(\varepsilon(p,T)-\mu)}
{T}\right]\right\}^{-1}.\eeq At $T\to 0$ one gets from Eqs. (1),
(3) the standard solution $n_F(p,T\to0)\to\theta(p_F-p)$, with
$\varepsilon(p\simeq p_F)-\mu=p_F(p-p_F)/M^*_L$, where $p_F$ is
the Fermi momentum, and $M^*_L$ is the Landau effective mass
\cite{lan}, \beq \frac{1}{M^*_L}
=\frac{1}{p}\frac{d\varepsilon(p,T=0)}{dp}|_{p=p_F}.\eeq It is
implied that $M^*_L$ is positive and finite at the Fermi momentum
$p_F$. As a result, the $T$-dependent corrections to $M^*_L$, to
the quasiparticle energy $\varepsilon (p)$, and other quantities,
start with $T^2$-terms. But this solution is not the only one
possible. There exist ``anomalous" solutions of Eq. (1) associated
with the so-called fermion condensation \cite{ks,ksk}. Being
continuous and satisfying the inequality $0<n(p)<1$ within some
region in $p$, such solutions $n(p)$ admit a finite limit for the
logarithm in Eq. (1) at $T\rightarrow 0$ yielding, \beq
\varepsilon(p)=\frac{\delta E[n(p)]} {\delta n(p)} =\mu; \:
p_i\leq p \leq p_f. \eeq At $T=0$ Eq. (5) determines
FCQPT, possessing solutions at some
$r_s=r_{FC}$ as soon as the effective inter-electron interaction
becomes sufficiently strong \cite{ksz}. From here on we shall call
a hole or electron system as electron one provided this will not
lead to confusion. In a simple electron
liquid, the effective inter-electron interaction is proportional to
the dimensionless average interparticle distance $r_s=r_0/a_B$,
with $r_0= \sqrt{2}/p_F$ being the average distance, and
$a_B$ is the Bohr radius. Equation (5) leads to the minimal value
of $E$ as a functional of $n(p)$ when in system under
consideration a strong rearrangement of the single particle
spectra can take place. We see from Eq. (5) that the occupation
numbers $n(p)$ become variational parameters: the solution $n(p)$
takes place if the energy $E$ can be lowered by alteration of the
occupation numbers. Thus, within the region $p_i<p<p_f$, the
solution $n(p)=n_F(p)+\delta n(p)$ deviates from the Fermi step
function $n_F(p)$ in such a way that the energy $\varepsilon(p)$
stays constant while outside this region $n(p)$ coincides with
$n_F(p)$. It is pertinent to note that the above general
consideration was verified by inspecting simple models. As the
result, it was shown that the onset of the FC does lead to lowering
the free energy \cite{ksk,dkss}.  It follows from the above
consideration that the superconductivity order parameter $\kappa({\bf
p})=\sqrt{n({\bf p})(1-n({\bf p}))}$ has a nonzero value over the
region occupied by FC. The superconducting gap $\Delta({\bf p})$
being linear in the coupling constant of the particle-particle
interaction $V_{pp}$ gives rise to the high value of $T_c$ because
one has $2T_c\simeq \Delta$ \cite{dkss} within the standard
Bardeen-Cooper-Schrieffer (BCS) theory \cite{bcs}. As it is shown in
Sec. III, if the superconducting gap $\Delta\neq 0$, the
quasiparticle effective mass becomes finite.  In consequence of these
features the density of states at the Fermi level becomes finite and
the involved quasiparticles are not localized. On the other hand,
even at $T=0$, $\Delta$ can vanish, provided $V_{pp}$ is repulsive or
absent.  Then, as it is seen from Eq. (5), the Landau quasiparticle
system becomes separated into two subsystems. The first contains the
Landau quasiparticles, while the second, related to FC, is localized
at the Fermi surface and formed by dispersionless quasiparticles. As
a result, the standard Kohn-Sham scheme for the single particle
equations is no longer valid beyond the point of the FC phase
transition \cite{vsl}. Such a behavior of systems with FC is clearly
different from what one expects from the well known local density
calculations. Therefore these calculations are hardly applicable to
describe systems with FC. It is also seen from Eq. (5) that a system
with FC has a well-defined Fermi surface.

Let us assume that FC has just taken place, that is $p_i\to p_f\to
p_F$, and the deviation $\delta n(p)$ is small. Expanding
functional $E[n(p)]$ in Taylor's series with respect to $\delta
n(p)$ and retaining the leading terms, one obtains from Eq. (5),
\beq \mu=\varepsilon({\bf p}) =\varepsilon_0({\bf p})+\int
F_L({\bf p},{\bf p}_1)\delta n({\bf p_1}) \frac{d{\bf
p}_1}{(2\pi)^2}; \: p_i\leq p \leq p_f,\eeq where $F_L({\bf
p},{\bf p}_1)=\delta^2 E/\delta n({\bf p})\delta n({\bf p}_1)$ is
the Landau interaction. Both the Landau interaction and the
single-particle energy $\varepsilon_0(p)$ are calculated at
$n(p)=n_F(p)$. It is seen from Eq. (6) that the FC quasiparticles
forms a collective state, since their energies are defined by the
macroscopical number of quasiparticles within the region
$p_i-p_f$, and vice versa. The shape of the spectra is not
effected by the Landau interaction, which, generally speaking,
depends on the system's properties, including the collective
states, impurities, etc. The only thing defined by the interaction
is the width of the region $p_i-p_f$, provided the interaction is
sufficiently strong to produce the FC phase transition at all.
Thus, we can conclude that the spectra related to FC are of
universal form, being dependent, as we will see below, mainly on
temperature $T$, if $T>T_c$, or on the superconducting gap at
$T<T_c$.

According to Eq. (1), the single-particle excitations within the
interval $p_i-p_f$ have at $T_c\leq T\ll T_f$ the shape
$\varepsilon(p,T)$ linear in T \cite{dkss,kcs}, which can be
simplified at the Fermi level, \beq
\varepsilon(p,T)-\mu(T)=T\ln\frac{1-n(p)}{n(p)} \simeq
T\frac{1-2n(p)}{n(p)}|_{p\simeq p_F}. \eeq $T_f$ is the
temperature, above which FC effects become insignificant
\cite{dkss}, \beq
\frac{T_f}{\varepsilon_F}\sim\frac{p_f^2-p_i^2}{2M\varepsilon_F}
\sim\frac{\Omega_{FC}}{\Omega_F}.\eeq Here $\Omega_{FC}$ is the FC
volume, $\varepsilon_F$ is the Fermi energy, and $\Omega_F$ is the
volume of the Fermi sphere. We note that at $T_c\leq T\ll T_f$ the
occupation numbers $n(p)$ are approximately independent of $T$,
being given by Eq. (5). One can imagine that at these temperatures
dispersionless plateau $\varepsilon(p)=\mu$ given by Eq. (5) is
slightly turned counter-clockwise about $\mu$. As a result, the
plateau is just a little tilted and rounded off at the end points.
According to Eq. (7) the effective mass $M^*_{FC}$ related to FC
is given by,
\beq M^*_{FC}\simeq
p_F\frac{p_f-p_i}{4T}.\eeq
To obtain Eq. (9) an approximation for the
derivative $dn(p)/dp\simeq -1/(p_f-p_i)$ was used. Having in mind
that $p_f-p_i\ll p_F$, and using (8) and (9) the following
estimates for the effective mass $M^*_{FC}$ are obtained, \beq
\frac{M^*_{FC}}{M_0} \sim
\frac{N(0)}{N_0(0)}\sim\frac{T_f}{T}.\eeq Eqs. (9) and (10) show
the temperature dependence of $M^*_{FC}$. In (10) $M_0$ denotes
the bare electron mass, $N_0(0)$ is the density of states of
noninteracting electron gas, and $N(0)$ is the density of states
at the Fermi level. Multiplying both sides of Eq. (9) by $p_f-p_i$
we obtain the energy scale $E_0$ separating the slow dispersing
low energy part, related to the effective mass $M^*_{FC}$, from
the faster dispersing relatively high energy part, defined by the
effective mass $M^*_{L}$ \cite{ms,ars}, \beq E_0\simeq 4T.\eeq It is
seen from Eq. (11) that the scale $E_0$ does not depend on the
condensate volume. The single particle excitations are defined
according to Eqs. (7) and (9) by the temperature and by $n(p)$,
given by Eq. (5). Thus, we are led to the conclusion that the
one-electron spectrum is negligible disturbed by thermal
excitations, impurities, etc, so that one observes the features of
the quantum protectorate \cite{rlp,pa}.

It is seen from Eq. (5) that at the point of FC phase transition
$p_f\to p_i\to p_F$, $M^*_{FC}$ and the density of states, as it
follows from Eqs. (5), (10), tend to infinity. One can conclude
that at $T=0$ and as soon as $r_s\to r_{FC}$, FCQPT takes place being
connected to the absolute growth of $M^*_{L}$. It is essential to
have in mind, that the onset of the charge density wave
instability in a many-electron system, such as electron liquid,
which takes place as soon as the effective inter electron constant
reaches its critical value $r_s=r_{cdw}$, is preceded
by the unlimited growth of the effective mass. Therefore, the FC
occurs before the onset of the charge density wave. Hence, at
$T=0$, when $r_s$ reaches its critical value $r_{FC}<r_{cdw}$, the
FCQPT inevitably takes place \cite{ksz}.
It is pertinent to note that this growth of the effective mass
with decreasing electron density was observed experimentally in a
metallic 2D electron system in silicon at $r_s\simeq 7.5$
\cite{skdk}. Therefore we can take $r_{FC}\sim 7.5$.
On the other hand, there exist charge density waves or
strong fluctuations of charge ordering in underdoped
high-$T_c$ superconductors \cite{grun}. Thus, the
formation of FC in high-$T_c$ compounds can be thought as a general
property of an electron liquid of low density which is embedded in
these solids, rather then an uncommon and anomalous solution of Eq.
(1) \cite{ksz}.  Beyond the point of FCQPT the condensate volume is
proportional to $(r_s-r_{FC})$ as well as $T_f/\varepsilon_F\sim
(r_s-r_{FC})/r_{FC}$ at least when $(r_s-r_{FC})/r_{FC}\ll 1$. Note,
that such a behavior is in accordance with the general properties of
second order phase transitions. Therefore, we can accept a model
relating systems with FC to high-$T_c$ compounds, assuming that the
effective coupling constant $r_s$ increases with decreasing doping,
exceeding its critical value $r_{FC}$ at the levels corresponding to
optimal doped samples.  We remark, that this critical value $r_{FC}$
corresponds to the $r_s$ values of highly overdoped samples
\cite{ksz}. As the result,
our quite natural model suggests that both quantities, $T_f$ and
condensate volume $\Omega_{FC}$, increase with decrease of doping.
Thus, these values are higher in underdoped samples as compared to
overdoped ones provided $r_s$ meets the mentioned above conditions.
While, in the highly overdoped regime only slight deviations from the
normal Fermi liquid are observed \cite{val1}. All these peculiar
properties are naturally explained within a model proposed in
\cite{ms,ars,ams,ms1} and allow to relate the doping level
$x$ regarded as the density of mobile charge
carriers (holes or electrons) per unit area to the density of Fermi
liquid with FC. We assume that $x_{FC}$ corresponds to the highly
overdoped regime at which FCQPT takes place, and introduce the
effective coupling constant $g_{eff}\sim (x-x_{FC})/x_{FC}$.  In our
model, the doping level $x$ at $x\leq x_{FC}$ in metals is related to
$(p_f-p_i)$ in the following way:  \beq g_{eff}\sim
\frac{(x_{FC}-x)}{x_{FC}}\sim \frac{(p_f-p_i)(p_f+p_i)}{p^2_F}
\sim \frac{p_f-p_i}{p_F}.\eeq
According to
experimental facts the large density of states at the Fermi level
reaches its maximum in the vicinity of the Hove singularities, that
is around the point $(\pi,0)$ of the Brillouin zone, or $\bar{M}$, in
high-$T_c$ compounds. The density of states reaches its minimum value
at the intersection of the so called nodal direction of the Brillouin
zone with the Fermi surface (see, e.g., \cite{shen}). The FC sets in
around the van Hove singularities \cite{kcs}, causing, according to
Eqs. (9) and (10), large density of states and large value of the
difference $(p_f-p_i)$ at the point $\bar{M}$. Then, the volume
$\Omega_{FC}$ and difference $(p_f-p_i)$ start to depend on the point
of the Fermi surface, say, on the angle $\phi$ along the Fermi
surface, which we count from the point $\bar{M}$ to the nodal
direction. Nonetheless, as it is seen from Eq. (11), $E_0$ remains
constant, being independent of the angle. It is not the case for the
effective mass $M^*_{FC}$, that can strongly depend upon the angle
via the difference $(p_f(\phi)-p_i(\phi))$ increasing from the nodal
direction towards $\bar{M}$, as it follows from Eq. (9). It is
pertinent to note that outside the FC region the single particle
spectrum is negligible affected by the temperature, being defined by
$M^*_L$. Thus, we come to the conclusion that a system with FC is
characterized by two effective masses: $M^*_{FC}$ that is related to
the single particle spectrum at lower energy scale, and $M^*_L$
describing the spectrum at higher energy scale.  These two effective
masses manifest itself as a break in the quasiparticle dispersion,
which can be approximated by two straight lines intersecting at the
energy $E_0$.  This break takes place at temperatures $T_c\leq T\ll
T_f$ in accordance with the experimental findings \cite{blk}, and, as
we will see, at $T\leq T_c$ corresponding to the experimental facts
\cite{blk,krc}. As to the quasiparticle formalism, it is applicable
to this problem since the width $\gamma$ of single particle
excitations is not large compared to their energy being proportional
$\gamma\sim T$ at $T>T_c$ \cite{dkss}. The lineshape can be
approximated by a simple Lorentzian \cite{ars}, being in accordance
with experimental data obtained from scans at a constant binding
energy \cite{vall}, see Sec. IV. Then, FC serves as a stimulating
source of new phase transitions which lift the degeneration of the
spectrum. For example, FC can generate the spin density wave, or
antiferromagnetic phase transition, thus leading to a whole variety
of the system's properties. Then, the onset of the charge density
wave is preceded by FCQPT, and both of these phases can coexist at
the sufficiently low density when $r_s\geq r_{cdw}$.  The simple
consideration presented above explains extremely large variety of
properties of high-$T_c$ superconductors.  We have seen above that
the superconductivity is strongly aided by FC, because both of the
phases are characterized by the same order parameter. As a result,
the superconductivity, removing the spectrum degeneration, ``wins the
competition" with the other phase transitions up to the critical
temperature $T_c$. We turn now to a consideration of both the
superconducting state and quasiparticle dispersions  at $T\leq T_c$.

\section{THE SUPERCONDUCTING STATE}

The explanation of the large values of the critical
temperature $T_c$, of the maximum value of the superconducting gap
$\Delta_1$, of the relation between $\Delta_1$ and the temperature
$T^*$ at which the pseudogap vanishes are, as years before, among the
main problems in the physics of high-temperature superconductivity. To
solve them, one needs to know the single-particle spectra of
corresponding metals. Recent studies of photoemission spectra
in copper oxide based compounds
discovered an energy scale in the spectrum of low-energy electrons in
copper oxides, which manifests itself as a kink in the
single-particle spectra \cite{blk,krc,vall,lanz}. As a result, the
spectra in the energy range (-200---0) meV can be described by two
straight lines intersecting at the binding energy
$E_0\sim(50-70)$ meV \cite{blk,krc}.
The existence of the energy scale $E_0$ could be attributed
to the interaction between electrons and the collective
excitations, for instance, phonons \cite{lanz}. On the other hand,
the analysis of the experimental data on the single-particle electron
spectra demonstrates that the perturbation of the spectra by phonons or
other collective states is in fact very small, therefore, the
corresponding state of electrons has to be described as a strongly
collectivized quantum state and was named ``quantum protectorate''
\cite{rlp,pa}. Thus, the interpretation of the above mentioned kink
as a consequence of electron-phonon interaction can very likely
be in contradiction with the quantum protectorate concept. To describe
the large values of $T_c$, the single-particle spectra and the kink,
the assumption can be used that the electron system of high-$T_c$
superconductor has undergone FCQPT.

The compounds are extremely complex materials having
a great number of competing degrees of freedom which produce
a great variety of physical properties. In turn, these properties
can compete and coexist with the superconductivity hindering the
understanding of the universal cause of the superconductivity. As a
result, it was suggested that the unique superconducting properties
in these compounds are defined by the presence of the Cu-O planes, by
the $d$-wave pairing symmetry, and by the existence of the pseudogap
phenomena in optimally doped and underdoped cuprates, see e.g.
\cite{tim,tk,vn}. However, recent studies of quasiparticle tunneling
spectra of cuprates have revealed that the pairing symmetry may change
from the $d$ to $s$-wave symmetry, depending on the hole, or
electron, doping level \cite{skin,bis,skin1}. Then, the high
temperature $s$-wave superconductivity has been observed in electron
doped infinite layer cuprates \cite{skin2} with a sharp
transition at $T=43$ K and the absence of pseudogap
\cite{chen}. Therefore, we can
conclude that the $d$-wave symmetry and the pseudogap phenomena are
not integral parts of . After all,
recent studies of high-$T_c$ superconductivity in C$_{60}$ crystals
\cite{skb} with the use of field-induced doping, when an increase in
the superconducting transition temperature $T_c$ to 52 K was achieved
in hole doped samples, have shown that the presence of the Cu-O
planes is not the necessary condition to observe high-$T_c$
superconductivity. Then, in lattice expanded C$_{60}$ by
intercalating CHCl$_3$ and CHBr$_3$ into the lattice, the higher
$T_c$ of 80 K in hole doped C$_{60}$/CHCl$_3$ and of 117 K in
C$_{60}$/CHBr$_3$ were observed \cite{skb1}. In the electron-doped
case, $T_c=11$ K is reached for C$_{60}$ crystals, $T_c=18$ K and
$T_c=26$ K were observed in samples intercalated with CHCl$_3$ and
CHBr$_3$ respectively \cite{skb1,skb2}. The described above
technique, when the corresponding dopant densities $x$ of electrons
or holes are induced by gate doping in a field-effect transistor
geometry, permits constructing the variation in
$T_c^{\alpha\gamma}(x)$ as a function of $x$ in a wide region of the
doping variation \cite{skb,skb1,skb2}. Here $\alpha$ denotes the
material, say C$_{60}$ or intercalated C$_{60}$, etc., and $\gamma$
denotes hole or electron doping. One important point to remember is
that in case of electron doped metals we have to treat $x$ as the
density of the mobile charge carriers, which corresponds to a narrow
variation region of $x$ around half-filling of the conduction band.
While the limitation of the hole doping variation is defined by the
electric breakdown strength of the gate oxide taking place at a
sufficiently high level of the doping \cite{skb1}. This technique
allows the study of properties of metals in question as a function of
the doping level $x$ without inducing disorder or possible defects
which could have a strong impact on the superconductivity. It is very
essential to note that the shape of the functions
$T_c^{\alpha\gamma}(x)$  is similar in samples with or without
intercalation, that is the shape does not depend on both $\alpha$ and
$\gamma$ \cite{skb1}. Moreover, this observation is also valid in the
case of the field induced superconductivity in both a spin-ladder
cuprate [CaCu$_2$O$_3$]$_4$ \cite{schs} and  CaCuO$_2$ \cite{schn}.
Thus, we can use a simple approximation \beq T_c^{\alpha\gamma}(x)
=T_1^{\alpha}T_2^{\gamma}(x_1-x)x,\eeq where the
coefficients $T_1^{\alpha}$ and $T_2^{\gamma}$ define the transition
temperature $T_c$ for a given hole (or electron) metal, and $x$ is
the density of the mobile charge carriers with $x$ is
obviously tending continuously to zero at the insulator-metal
transition. It is directly follows from Eq. (13) that the transition
temperature reaches its maximum value $T^M_c$ at the optimal doping
level $x_{opt}$ \beq
T^M_c=T_c^{\alpha\gamma}(x_{opt})=T_1^{\alpha}T_2^{\gamma}
\left(\frac{x_1}{2}\right)^2.\eeq
Now we can calculate the value of $x_{opt}$ for the different
hole and electron metals studied in \cite{skb,skb1,skb2,schs,schn} in
terms of the dimensionless parameter $r_s$, $\pi r_s^2a^2_B=1/x$.
We have that $r_s^{opt}\sim
10$ and, thus, $r_s^{opt}$ is approximately independent of the metals.
As a result we can recognize that these striking experimental facts,
the general shape of the function $T_c(x)$ and the constant value of
$r_s^{opt}$, point to a fact, that the generic properties of
high-$T_c$ superconductivity are defined by the 2D charge (electron
or hole) strongly correlated liquid rather then by solids which hold
this liquid. While the solids arrange the presence of the pseudogap
phenomena, the $s$ or $d$-wave pairing symmetry, the electron-phonon
coupling constant defining $T_c$, the variation region of $x$, and so
on.

At $T=0$, the ground state energy
$E_{gs}[\kappa({\bf p}),n({\bf p})]$ of 2D electron
liquid is a functional of the order parameter of the superconducting
state $\kappa({\bf p})$ and of the quasiparticle occupation numbers
$n({\bf p})$ and is determined by the known equation of the
weak-coupling theory of superconductivity, see e.g. \cite{til}
\beq E_{gs}=E[n({\bf p})]
+\int \lambda_0V({\bf p}_1,{\bf
p}_2)\kappa({\bf p}_1) \kappa^*({\bf p}_2) \frac{d{\bf p}_1d{\bf
p}_2}{(2\pi)^4}.\eeq Here  $E[n({\bf p})]$ is the ground-state energy
of normal Fermi liquid, $n({\bf p})=v^2({\bf
p})$ and $\kappa({\bf p})=v({\bf p})\sqrt{1-v^2({\bf p})}$.
It is assumed that the pairing interaction $\lambda_0V({\bf
p}_1,{\bf p}_2)$ is weak. Minimizing $E_{gs}$ with
respect to $\kappa({\bf p})$
we obtain the equation connecting the single-particle energy
$\varepsilon({\bf p})$ to $\Delta({\bf p})$,
\beq \varepsilon({\bf p})-\mu=\Delta({\bf p})
\frac{1-2v^2({\bf p})} {2\kappa({\bf p})},\eeq
here the single-particle energy $\varepsilon({\bf p})$ is
determined by the Landau equation (2).
The equation for the
superconducting gap $\Delta({\bf p})$
takes form
\beq \Delta({\bf p})
=-\int\lambda_0V({\bf p},{\bf p}_1)\kappa({\bf p}_1)
\frac{d{\bf p}_1}{4\pi^2}
=-\frac{1}{2}\int\lambda_0
V({\bf p},{\bf p}_1) \frac{\Delta({\bf p}_1)}
{\sqrt{(\varepsilon({\bf p}_1)-\mu)^2+\Delta^2({\bf p}_1)}}
\frac{d{\bf p}_1}{4\pi^2}.\eeq
If $\lambda_0\to 0$, then, the maximum value $\Delta_1\to 0$, and
Eq. (16) reduces to Eq. (5)
\cite{ks} \beq \varepsilon({\bf
p})-\mu=0,\: {\mathrm {if}}\,\,\, 0<n({\bf p})<1;\: p_i\leq p\leq
p_f.\eeq
Now we can study relationships between the state defined by Eq. (18)
and the superconductivity.
At $T=0$, Eq. (18) defines a particular state of Fermi liquid
with FC for which the modulus of the order
parameter $|\kappa({\bf p})|$ has finite values in the $L_{FC}$
range of momenta $p_i\leq p\leq p_f$, and $\Delta_1\to 0$ in the
$L_{FC}$. Such a state can be considered as superconducting,
with infinitely small value of $\Delta_1$ so that the
entropy of this state is equal to zero. It is obvious, that
this state, being driven by the quantum
phase transition, disappears at $T>0$ \cite{ms}. When $p_i\to
p_F\to p_f$, Eq. (18) determines the point $r_{FC}$ at which the FCQPT
takes place. It follows from Eq. (18) that the system brakes into
two quasiparticle subsystems: the first subsystem in the $L_{FC}$ range
is occupied by the quasiparticles with the effective mass
$M^*_{FC}\propto1/\Delta_1$,
while the second one is occupied by quasiparticles with finite
mass $M^*_L$ and momenta $p<p_i$. If $\lambda_0\neq0$, $\Delta_1$
becomes finite, leading to a finite value of the effective mass
$M^*_{FC}$ in $L_{FC}$, which can be obtained from Eq. (16)
\cite{ms,ars}
\beq M^*_{FC} \simeq p_F\frac{p_f-p_i}{2\Delta_1}.\eeq
As to the energy scale, it is determined by the parameter $E_0$:
\beq E_0=\varepsilon({\bf p}_f)-\varepsilon({\bf p}_i)
\simeq2\frac{(p_f-p_F)p_F}{M^*_{FC}}\simeq 2\Delta_1.\eeq

It is reasonably safe
to suggest that we have come back to the Landau theory by integrating
out high energy degrees of freedom and introducing the
quasiparticles. The sole difference between the Landau Fermi liquid
and Fermi liquid undergone FCQPT is that we have to expand the
number of relevant low energy degrees of freedom by adding both a new
type of quasiparticles with the effective mass $M^*_{FC}$, given by
Eq. (19), and the energy scale $E_0$ given by Eq. (20). We have also
to bear in mind that the properties of these new quasiparticles of a
Fermi liquid with FC cannot be separated from the properties of the
superconducting state, as it follows from Eqs. (16), (19) and (20).
We may say that the quasiparticle system in the range $L_{FC}$
becomes very ``soft'' and is to be considered as a strongly correlated
liquid.  On the other hand, the system's properties and dynamics are
dominated by a strong collective effect having its origin in FCQPT
and determined by the macroscopic number of quasiparticles in the
range $L_{FC}$.  Such a system cannot be disturbed by the scattering
of individual quasiparticles and has features of a quantum
protectorate \cite{ms,rlp,pa}.

We assume that the range $L_{FC}$ is small, $(p_f-p_F)/p_F\ll1$, and
$2\Delta_1\ll T_f$ so that the order parameter $\kappa({\bf p})$ is
governed mainly by the FC \cite{ms,ams}. To solve Eq. (17)
analytically, we take the Bardeen-Cooper-Schrieffer (BCS)
approximation for the interaction \cite{bcs}:  $\lambda_0V({\bf
p},{\bf p}_1)=-\lambda_0$ if $|\varepsilon({\bf p})-\mu|\leq
\omega_D$, the interaction is zero outside this region, with
$\omega_D$ being the characteristic phonon energy.  As a result, the
gap becomes dependent only on the temperature, $\Delta({\bf
p})=\Delta_1(T)$, being independent of the momentum, and Eq. (17)
takes the form \beq 1=N_{FC}\lambda_0\int_0^{E_0/2}\frac{d\xi}
{\sqrt{\xi^2+\Delta^2_1(0)}}
+N_{L}\lambda_0\int_{E_0/2}^{\omega_D}\frac{d\xi}
{\sqrt{\xi^2+\Delta^2_1(0)}}.\eeq
Here we set $\xi=\varepsilon({\bf p})-\mu$ and introduce the density
of states $N_{FC}$ in the $L_{FC}$, or $E_0$, range. As
it follows from Eq. (19), $N_{FC}=(p_f-p_F)p_F/2\pi\Delta_1(0)$.
The density of states $N_{L}$ in the range
$(\omega_D-E_0/2)$ has the standard form $N_{L}=M^*_{L}/2\pi$.
If the energy scale $E_0\to 0$, Eq. (21) reduces to the BCS equation.
On the other hand, assuming that $E_0\leq2\omega_D$ and omitting the
second integral in the right hand side of Eq. (21), we obtain
\beq\Delta_1(0)=\frac{\lambda_0 p_F(p_f-p_F)}{2\pi}\ln(1+\sqrt{2})=
2\beta\varepsilon_F\frac{p_f-p_F}{p_F}\ln(1+\sqrt{2}),\eeq
where the Fermi energy $\varepsilon_F=p_F^2/2M^*_L$, and
dimensionless coupling constant $\beta=\lambda_0 M^*_L/2\pi$.
Taking the usual values of the dimensionless coupling constant
$\beta\simeq 0.3$, and $(p_f-p_F)/p_F\simeq 0.2$, we
get from Eq. (22) the large value of $\Delta_1(0)\sim
0.1\varepsilon_F$, while for normal metals one has $\Delta_1(0)\sim
10^{-3}\varepsilon_F$.  Taking into account the omitted integral, we
obtain \beq\Delta_1(0)\simeq 2\beta\varepsilon_F
\frac{p_f-p_F}{p_F}\ln(1+\sqrt{2})\left(1+\beta
\ln\frac{2\omega_D}{E_0}\right).\eeq
It is seen from Eq. (23) that the correction due to the second
integral is small, provided $E_0\simeq2\omega_D$. Below we show
that $2T_c\simeq \Delta_1(0)$, which leads to the conclusion that
there is no isotope effect since $\Delta_1$ is independent of
$\omega_D$. But this effect is restored as $E_0\to 0$.
Assuming $E_0\sim\omega_D$ and $E_0>\omega_D$, we see that Eq. (21)
has no standard solutions $\Delta(p)=\Delta_1(T=0)$ because
$\omega_D<\varepsilon(p\simeq p_f)-\mu$ and the interaction vanishes at
these momenta. The only way to obtain solutions is to restore the
condition $E_0<\omega_D$. For instance, we can define the momentum
$p_D<p_f$ such that \beq \Delta_1(0)=2\beta\varepsilon_F
\frac{p_D-p_F}{p_F}\ln(1+\sqrt{2})=\omega_D,\eeq
while the other part in the $L_{FC}$ range can be occupied by a
gap $\Delta_2$ of the different sign, $\Delta_1(0)/\Delta_2<0$. It
follows from Eq. (24) that the isotope effect is presented, while
the both gaps can have $s$-wave symmetry.  A more detailed analysis
will be published elsewhere.

At $T\simeq T_c$, Eqs. (19) and (20) are replaced by
the equation, which is valid also at $T_c\leq T\ll T_f$
in accord with Eq. (9) \cite{ms}
\beq M^*_{FC}\simeq p_F\frac{p_f-p_i}{4T_c},\,\,\,
E_0\simeq 4T_c;\,\,\,{\mathrm {if}}\,\,\,T_c\leq T:
\,\,\, M^*_{FC}\simeq p_F\frac{p_f-p_i}{4T},\,\,\,
E_0\simeq 4T.\eeq
Equation (21) is replaced by its conventional finite temperature
generalization
\beq
1=N_{FC}\lambda_0\int_0^{E_0/2}
\frac{d\xi}
{\sqrt{\xi^2+\Delta^2_1(T)}}
\tanh\frac{\sqrt{\xi^2+\Delta^2_1(T)}}{2T}
+ N_{L}\lambda_0\int_{E_0/2}^{\omega_D}
\frac{d\xi}
{\sqrt{\xi^2+\Delta^2_1(T)}}
\tanh\frac{\sqrt{\xi^2+\Delta^2_1(T)}}{2T}.\eeq
Putting $\Delta_1(T\to T_c)\to 0$, we obtain from Eq. (26)
\beq 2T_c\simeq \Delta_1(0),\eeq
with $\Delta_1(T=0)$ being given by Eq. (21).
By comparing Eqs. (19), (25) and (27), we see that $M^*_{FC}$ and
$E_0$ are almost temperature independent at $T\leq
T_c$. In the same way, as it was done in Sec. II, we can conclude
that $E_0$ does not change along the Fermi surface, while
$M^*_{FC}$ increases when moving from the nodal direction to the point
$\bar{M}$. Now a few remarks are in order. One can define $T_c$ as
the temperature when $\Delta_1(T_c)\equiv 0$. At $T\geq T_c$, Eq.
(26) has only the trivial solution $\Delta_1\equiv 0$. On the other
hand, $T_c$ can be defined as a temperature at which the
superconductivity vanishes. Thus, we deal with two different
definitions, which can lead to two different temperatures $T_c$ and
$T^*$ in case of the $d$-wave symmetry of the gap. It was shown
\cite{ars,sh} that in the case of the d-wave superconductivity,
taking place in the presence of the FC, there exist a nontrivial
solutions of Eq. (26) at $T_c\leq T\leq T^*$ corresponding to the
pseudogap state. It happens when the gap occupies only such a part of
the Fermi surface, which shrinks as the temperature increases. Here
$T^*$ defines the temperature at which $\Delta_1(T^*)\equiv0$ and the
pseudogap state vanishes. The superconductivity is destroyed at
$T_c$, and the ratio $2\Delta_1/T_c$ can vary in a wide range and
strongly depends upon the material's properties, as it follows from
consideration given in \cite{ars,ms1,sh}. Therefore, provided there
exists the pseudogap above $T_c$, then $T_c$ is to be replaced by
$T^*$, and Eq. (27) takes the form \beq 2T^*\simeq \Delta_1(0).\eeq
The ratio $2\Delta_1/T_c$ can reach very high values. For instance,
in the case of Bi$_2$Sr$_2$CaCu$_2$Q$_{6+\delta}$, where the
superconductivity and the pseudogap are considered to be of the
common origin, $2\Delta_1/T_c$ is about 28, while the ratio
$2\Delta_1/T^*\simeq 4$, which is also valid for various cuprates
\cite{kug}. Thus, Eq. (28) gives good description of the experimental
data. We remark that Eq.  (21) gives also good description of the
maximum gap $\Delta_1$ in the case of the d-wave superconductivity,
because the different regions with the maximum absolute value of
$\Delta_1$ and the maximal density of states can be considered as
disconnected \cite{abr}. Therefore, the gap in this region is formed
by attractive phonon interaction which is approximately independent
of the momenta.

Consider now two possible types of the superconducting gap
$\Delta({\bf p})$ given by Eq. (17) and defined by interaction
$\lambda_0V({\bf p},{\bf p}_1)$. If this interaction is dominated by
a phonon-mediated attraction, the even solution of Eq. (17) with the
$s$-wave, or the $s+d$ mixed waves, will have the lowest energy.
Provided the pairing interaction $\lambda_0V({\bf p}_1,{\bf p}_2)$ is
the combination of both the attractive interaction and
sufficiently strong repulsive interaction, the $d$-wave odd
superconductivity can take place, see e.g. \cite{abr}.  But both the
$s$-wave even symmetry and $d$-wave odd one lead to the approximately
same value of the gap $\Delta_1$ in Eq.  (21) \cite{ams}.  Therefore,
the non-universal pairing symmetries in high-$T_c$ superconductivity
is likely the result of the pairing interaction, and the $d$-wave
pairing symmetry cannot be considered as essential to high-$T_c$ in
keeping with experimental findings \cite{skin,bis,skin1,skin2,chen}.
In case, if there were only the $d$-wave pairing, the crossover from
superconducting gap to pseudogap can take place, so that the
superconductivity is destroyed at the temperature $T_c$, with the
superconducting gap being smoothly transformed into the pseudogap
which closes at some temperature $T^*>T_c$ \cite{ms1,sh}. In the case
of the $s$-wave pairing we can expect the absence of the pseudogap
phenomena in accordance with the experimental observation, see
\cite{chen} and references therein.

We turn now to a consideration of the maximum value of the
superconducting gap $\Delta_1$ as a function of the density $x$
of mobile charge carriers. Being rewritten in terms of $x$
and $x_{FC}$ related to the variables $p_i$ and $p_f$ by Eq. (12),
Eq. (22) takes the form \beq\Delta_1\propto\beta(x_{FC}-x)x.\eeq
Here we take into account that the Fermi level $\varepsilon_F\propto
p_F^2$, the density $x\propto p_F^2$, and thus,
$\varepsilon_F\propto x$.
Considering the field induced superconductivity, we can safely assume
that $T_c\propto\Delta_1$ because this technique allows the study of
properties of metal in question as a function of $x$ without
inducing additional defects or disorder which can have a dramatic
impact on the transition temperature. Then, instead of Eq. (29) we
have \beq T^{\alpha\gamma}_c
\propto\beta^{\alpha}\beta^{\gamma}(x_{FC}-x)x.\eeq
In Eq. (30), we made the natural change
$\beta=\beta^{\alpha}\beta^{\gamma}$ since the coupling constant
$\beta$ is fixed by the properties of metal in question.
Following reference \cite{bkr}, we
take that hole doped metals differ from electron doped ones only in
the magnitude of the coupling constant $\beta^{\gamma}$ which is
smaller in case of electron doped metals.
Now it is seen that Eq. (30) coincides with Eq. (13)
producing the universal optimal doping level
$x_{opt}=x_{FC}/2=x_1/2$
in line with the experimental facts.
In our model, we have
$x_{opt}/x_{FC}=(r_s^{opt}/r_{FC})^2=2$, taking the value
$r_s^{opt}\simeq 10$, we obtain $r_{FC}\simeq 7.0$. This result is
in a reasonable agreement with the experimental value $r_{FC}\sim
7.5$ corresponding to sharp increase of the effective mass
\cite{skdk}. In line with facts
\cite{skb1}, it follows from Eq. (30) that among the hole doped
fullerides, the $T_c$ ratios for
C$_{60}$/CHBr$_{3}$-C$_{60}$/CHCl$_{3}$-C$_{60}$ have to be the same
as in the case of the respective electron doped fullerides because
the factor $\beta^{\gamma}$ drops out of the ratios.

As an example of the implementation of the previous analysis let us
consider the main features of a room-temperature superconductor.
The superconductor has to be a quasi two-dimensional
structure, presented by infinite-layer compounds or by
field-induced superconductivity in gated structures. As it follows
from Eq. (22), $\Delta_1\sim \beta \varepsilon_F\propto \beta/r_s^2$.
Noting that FCQPT takes place in 3D systems at $r_s\sim 20$ and in 2D
systems at $r_s\sim 8$ \cite{ksz}, we can expect that
$\Delta_1$ of 3D system comprises 10\% of the corresponding maximum
value of 2D superconducting gap, reaching values as high as 60 meV
for underdoped crystals with $T_c=70$ \cite{mzo}.
On the other hand, it is seen form Eq. (22), that $\Delta_1$ can be
even large, $\Delta_1\sim 75$ meV, and one can expect $T_c\sim 300$
K in the case of the $s$ wave pairing as it follows from the simple
relation $2T_c\simeq \Delta_1$.  In fact, we can safely take
$\varepsilon_F\sim 300$ meV, $\beta\sim 0.5$ and
$(p_f-p_i)/p_F\sim0.5$. Thus, we can conclude that a possible
room-temperature superconductor has to be the $s$-wave superconductor
in order to get rid of the pseudogap phenomena, which tremendously
reduces the transition temperature. The density $x$ of the mobile
charge carriers must satisfy the condition $x\leq x_{FC}$ and be
flexible to reach the optimal doping level. It is worth
noting that the coupling constant $\beta$ has to be sufficiently big
because FC giving rise to the order parameter $\kappa({\bf p})$ does
not produce the gap $\Delta$ by itself. For instance, the coupling
constant can be enhanced by an intercalation as it is done
for fullerides \cite{skb1,bkr}.

Now we turn to the calculations of the gap and the specific heat at
the temperatures $T\to T_c$. It is worth noting that this
consideration is valid provided $T^*=T_c$, otherwise the considered
below discontinuity is smoothed out over the temperature range
$T^{*}\div T_c$. For the sake of simplicity, we calculate the main
contribution to the gap and the specific heat coming from the FC.
The function $\Delta_1(T\to T_c)$ is found from Eq. (26)
upon expanding the right hand side of the first integral in powers
of $\Delta_1$ and omitting the contribution from the
second integral on the right hand side of Eq. (26). This procedure
leads to the following equation \cite{ams}
\beq \Delta_1(T)\simeq
3.4T_c\sqrt{1-\frac{T}{T_c}}.\eeq
Thus, the gap in the spectrum of the single-particle excitations has
quite usual behavior. To calculate the specific heat, the
conventional expression for the entropy $S$ \cite{bcs} can be used
\beq S=2\int\left[f({\bf p})\ln f({\bf p})
+(1-f({\bf p}))\ln(1-f({\bf p}))\right]\frac{d{\bf p}}{(2\pi)^2},\eeq
where
\beq f({\bf p})=\frac{1}{1+\exp[
E({\bf p})/T]};\,\,\,
E({\bf p})
=\sqrt{(\varepsilon({\bf p})-\mu)^2+\Delta_1^2(T)}.\eeq
The specific heat $C$ is determined by
\begin{eqnarray}
C=T\frac{dS}{dT}&\simeq&4\frac{N_{FC}}{T^2}\int_0^{E_0}
f(E)(1-f(E))\left[E^2+T\Delta_1(T)
\frac{d\Delta_1(T)}{dT}\right]d\xi\nonumber\\
& &+4\frac{N_{L}}{T^2}\int_{E_0}^{\omega_D}
f(E)(1-f(E))\left[E^2+T\Delta_1(T)
\frac{d\Delta_1(T)}{dT}\right]d\xi.
\end{eqnarray}
When deriving Eq. (34) we again use the variable $\xi$ and the
densities of states $N_{FC}$, $N_{L}$, just as before in connection to
Eq. (21), and use the notation $E=\sqrt{\xi^2+\Delta_1^2(T)}$.
Equation (34) predicts the conventional discontinuity $\delta C$ in
the specific heat $C$ at $T_c$ because of the last term in the square
brackets of Eq. (34).  Upon using Eq. (31) to calculate this term
and omitting the second integral on the right hand side of Eq. (34),
we obtain \beq \delta C\simeq\frac{3}{2\pi}(p_f-p_i)p_F.\eeq In
contrast to the conventional result when the discontinuity is a
linear function of $T_c$, $\delta C$ is independent of the critical
temperature $T_c$ because the density of state varies inversely with
$T_c$ as it follows from Eq. (25). Note, that deriving Eq. (35) we
take into account the main contribution coming from the FC. This
contribution vanishes as soon as $E_0\to0$ and the second integral of
Eq. (34) gives the conventional result.

\section{THE LINESHAPE OF THE SINGLE-PARTICLE SPECTRA}

Consider the lineshape $L(q,\omega)$ of the single-particle
spectrum which is a function of two variables. Measurements
carried out at a fixed binding energy $\omega=\omega_0$, where
$\omega_0$ is the energy of a single-particle excitation, determine
the lineshape $L(q,\omega=\omega_0)$ as a function of the momentum $q$.
We have shown above that $M^*_{FC}$ is finite and constant at
$T\leq T_c$. Therefore, at excitation energies $\omega\leq E_0$ the
system behaves like an ordinary superconducting Fermi liquid with the
effective mass given by Eq. (19) \cite{ms,ars}. At $T_c\leq T$ the low
energy effective mass $M^*_{FC}$ is finite and is given by Eq. (9).
Once again, at the energies $\omega<E_0$, the system behaves as a
Fermi liquid, the single-particle spectrum is well defined, while the
width of single-particle excitations is of the order of $T$
\cite{ms,dkss}. This behavior was observed in experiments on
measuring the lineshape at a fixed energy \cite{vall}.
It is pertinent to note that recent measurements of the lineshape
suggest that quasiparticle excitation even in the $(\pi,0)$ region of
the Brillouin zone of Bi$_2$Sr$_2$CaCu$_2$Q$_{8+\delta}$ (Bi2212) are
much better defined then previously believed from earlier Bi2212 data
\cite{feng}. We remark that our model is in accordance with these
measurements suggesting that well-defined quasiparticles exist at the
Fermi level.

The lineshape can also be determined as a function
$L(q=q_0,\omega)$ at a fixed $q=q_0$.  At small $\omega$, the lineshape
resembles the one considered above, and $L(q=q_0,\omega)$ has a
characteristic maximum and width. At energies $\omega\geq E_0$,
quasiparticles with the mass $M^*_{L}$ come into play, leading to a
growth of the function $L(q=q_0,\omega)$. As a result, the function
$L(q=q_0,\omega)$ possesses the known peak-dip-hump structure
\cite{dess} directly defined by the existence of the two effective
masses $M^*_{FC}$ and $M^*_L$ \cite{ms,ars}. To have more
quantitative and analytical insight into the problem we use the
Kramers-Kr\"{o}nig transformation to construct the imaginary part
${\mathrm{Im}}\Sigma({\bf p},\varepsilon)$ of the self-energy
$\Sigma({\bf p},\varepsilon)$ starting with the real one
${\mathrm{Re}}\Sigma({\bf p},\varepsilon)$ which defines the effective
mass \cite{mig} \beq \frac{1}{M^*}=\left(\frac{1}{M}+\frac{1}{p_F}
\frac{\partial{\mathrm{Re}}\Sigma}{\partial p}\right)/
\left(1-\frac{\partial{\mathrm{Re}}\Sigma}{\partial
\varepsilon}\right).\eeq
Here $M$ is the bare mass, while the relevant momenta $p$ and
energies $\varepsilon$ are subjected to the conditions:
$|p-p_F|/p_F\ll 1$, and $\varepsilon/\varepsilon_F\ll 1$.
We take ${\mathrm{Re}}\Sigma({\bf p},\varepsilon)$ in the simplest form
which accounts for the change of the effective mass at
the energy scale $E_0$:
\beq {\mathrm{Re}}\Sigma({\bf p},\varepsilon)=-\varepsilon
\frac{M^*_{FC}}{M}+\left(\varepsilon-\frac{E_0}{2}\right)
\frac{M^*_{FC}-M^*_{L}}{M}\left[\theta(\varepsilon-E_0/2)
+\theta(-\varepsilon-E_0/2)\right].\eeq
Here $\theta(\varepsilon)$ is the step function. Note that in
order to ensure a smooth transition from the single-particle
spectrum characterized by $M^*_{FC}$ to the spectrum defined by
$M^*_{L}$ the step function is to be substituted by some smooth
function. Upon inserting
Eq. (37) into Eq. (36) we can check that inside the interval
$(-E_0/2,E_0/2)$ the effective mass $M^*\simeq M^*_{FC}$, and outside
the interval $M^*\simeq M^*_{L}$. By applying the Kramers-Kr\"{o}nig
transformation to ${\mathrm{Re}}\Sigma({\bf p},\varepsilon)$, we
obtain the imaginary part of the self-energy \cite{ams} \beq
{\mathrm{Im}}\Sigma({\bf p},\varepsilon)\sim
\varepsilon^2\frac{M^*_{FC}}{\varepsilon_F M}+
\frac{M^*_{FC}-M^*_L}{M}\left(
\varepsilon\ln\left|\frac{\varepsilon+E_0/2}
{\varepsilon-E_0/2}\right|+
\frac{E_0}{2}\ln\left|\frac{\varepsilon^2-E^2_0/4}
{E^2_0/4}\right|\right).\eeq
We can see from Eq. (38)
that at $\varepsilon/E_0\ll 1$ the imaginary part is proportional
to $\varepsilon^2$; at $2\varepsilon/E_0\simeq 1$
${\mathrm{Im}}\Sigma\sim \varepsilon$; at $E_0/\varepsilon\ll 1$
the main contribution to the imaginary part is approximately
constant. This is the behavior that gives rise to the known
peak-dip-hump structure. Then, it is seen from Eq. (38) that when
$E_0\to 0$ the second term on the right hand side tends to zero, the
single-particle excitations become  better defined resembling
that of a normal Fermi liquid, and the peak-dip-hump structure
eventually vanishes. On the other hand, the quasiparticle amplitude
$a({\bf p})$ is given by \cite{mig}
\beq \frac{1}{a({\bf p})}=1-\frac{\partial
{\mathrm{Re}}\Sigma({\bf p},\varepsilon)}{\partial\varepsilon}.\eeq
It follows from Eq. (36)
that the quasiparticle amplitude $a({\bf p})$ rises as the effective
mass $M^*_{FC}$ decreases.  Since, as it follows from
Eq. (12), $M^*_{FC}\sim (p_f-p_i)/p_F\sim (x_{FC}-x)/x_{FC}$, we are
led to a conclusion that the amplitude $a({\bf p})$ rises as the
doping level rises, and the single-particle excitations become better
defined in highly overdoped samples.
It is worth noting that such a behavior was
observed experimentally in so highly overdoped Bi2212 that the gap
size is about 10 meV \cite{val1}.  Such a small size of the gap
testifies that the region occupied by the FC is small since
$E_0/2\simeq \Delta_1$. Quasiparticles
located at the intersection of the nodal direction of the Brillouin
zone with the Fermi level should also be well-defined
comparatively to quasiparticles located at the point $\bar{M}$
because of the decrease of the effective mass $M^*_{FC}$ when moving
from the point $\bar{M}$ to the nodal direction, as it was discussed
in Sections II and III. This is especially true in regard to strongly
underdoped samples.

\section{SUMMARY}

In conclusion, we have shown that the theory of high temperature
superconductivity based on the fermion-condensation quantum phase
transition and on the conventional theory of superconductivity
permits to describe high values of $T_c$, $T^*$ and of the maximum
value of the gap $\Delta_1$, which may be as big as $\Delta_1\sim
0.1\varepsilon_F$ or even larger.  We have also traced the transition
from conventional superconductors to high-$T_c$ and demonstrated that
in the highly overdoped cuprates the single-particle excitations
become much better defined, resembling that of a normal Fermi liquid.

We have also shown by a simple, self-consistent
analysis, that the general features of the shape of the critical
temperature $T_c$ as a function of the density $x$ of the mobile
carriers in the metals and the value of the optimal doping $x_{opt}$
can be understood within the framework of the theory of the
high-$T_c$ superconductivity based on FCQPT. We have demonstrated
that neither the $d$-wave pairing symmetry, nor the pseudogap
phenomenon, nor the presence of the Cu-O planes  are of importance
for the existence of the high-temperature superconductivity.
As a result, we can conclude that the generic properties of
high-temperature superconductors are defined by the 2D charge
(electron or hole) strongly correlated liquid rather then by solids
which hold this liquid. While the solids arrange the presence of the
pseudogap phenomena, the $s$-wave pairing symmetry or $d$-wave one,
the electron-phonon coupling constant defining $T_c$, the variation
region of $x$, and so on. The main features of a room-temperature
superconductor have also been outlined.

This work was supported by the Russian Foundation for
Basic Research, project 01-02-17189.

\end{document}